\documentclass[prc,aps,twocolumn,showpacs]{revtex4}
\newcommand{\msun}{M$_{\odot}$ }
\usepackage{graphicx}
\begin{document}
\pagebreak
\title{Reevaluation of the $^{22}$Na(p,$\gamma$) reaction rate: Implications for the detection of $^{22}$Na gamma rays from novae}
\author{D.G.Jenkins\footnote{present address: Department of Physics, University of York, Heslington, York YO10 5DD, U.K.}}
\affiliation{Physics Division, Argonne National Laboratory, Argonne, IL 60439}
\affiliation{Department of Physics, University of Pennsylvania, Philadelphia, PA 19104}
\author{C.J.Lister}
\author{R.V.F.Janssens}
\author{T.L.Khoo}
\author{E.F.Moore}
\author{K.E.Rehm}
\author{B.Truett}
\author{A.H.Wuosmaa}
\affiliation{Physics Division, Argonne National Laboratory, Argonne, IL 60439}
\author{M.Freer}
\affiliation{School of Physics and Astronomy, University of Birmingham, Birmingham B15 2TT, U.K}
\author{B.R.Fulton}
\affiliation{Department of Physics, University of York, Heslington, York YO10 5DD, U.K.}
\author{J.Jos\'e}
\affiliation{Dept. F\'{\i}sica i Enginyeria Nuclear (UPC)}
\affiliation{Institut d'Estudis Espacials de Catalunya (IEEC), E-08034 Barcelona, Spain}

\date{\today}

\begin{abstract}

Understanding the processes which create and destroy $^{22}$Na is important for diagnosing classical nova outbursts. Conventional $^{22}$Na(p,$\gamma$) studies are complicated by the need to employ radioactive targets. In contrast, we have formed the particle-unbound states of interest through the heavy-ion fusion reaction, $^{12}$C($^{12}$C,n)$^{23}$Mg and used the Gammasphere array to investigate their radiative decay branches. Detailed spectroscopy was possible and the  $^{22}$Na(p,$\gamma$) reaction rate has been re-evaluated. New hydrodynamical calculations incorporating the upper and lower limits on the new rate suggest a reduction in the yield of $^{22}$Na with respect to previous estimates, implying a reduction in the maximum detectability distance for $^{22}$Na $\gamma$ rays from novae. 
  
\end{abstract} 


\maketitle

The nucleus $^{22}$Na is potentially of considerable interest in the diagnosis of
classical nova outbursts \cite{clayton74} -- violent events that take 
place on the white dwarf component of a close stellar binary system. This radioactive
isotope decays, with a 2.602 yr half-life, into a short-lived 
excited state of ${}^{22}$Ne, emitting a $1.275$ MeV $\gamma$ ray. 
Novae have been proposed as the principal galactic sites for the synthesis of $^{22}$Na. Explosions within
a few kiloparsecs of the Sun may provide detectable $\gamma$-ray fluxes
associated with $^{22}$Na decay (see Ref. \cite{hernanz2002} for a recent review). 
Several experimental searches for this $\gamma$-ray signature of classical novae
have been performed in the last twenty five years, including
balloon-borne experiments \cite{leventhal}, the OSSE and COMPTEL experiments
on board the CGRO \cite{leising93,iyudin95}, and the recently-launched INTEGRAL mission \cite{hernanz2002}.
 Constraints on the overall amount of $^{22}$Na ejected into the interstellar medium during 
nova outbursts have been derived from the earlier
experiments. The most restrictive ones correspond to 
measurements performed with COMPTEL of five recent neon-type novae \cite{iyudin95} which led to an upper limit 
of $3 \times 10^{-8}$ \msun of $^{22}$Na ejected by any nova in the Galactic disk.

The synthesis of $^{22}$Na in novae has been extensively investigated in the last two
decades (a recent review is found in Ref. \cite{jose2002}). 
Proton-capture reactions on the `seed' nuclei $^{20}$Ne are responsible for the
synthesis of significant amounts of the unstable nucleus $^{21}$Na. This is followed 
either by its $\beta^+$-decay into $^{21}$Ne (`cold' NeNa cycle), that leads to
$^{22}$Na by means of $^{21}$Ne(p,$\gamma$)$^{22}$Na, or, for high enough temperatures,
by additional proton-captures on $^{21}$Na (`hot' NeNa cycle) leading to $^{22}$Mg that 
subsequently decays into $^{22}$Na. 

The production of $^{22}$Na in novae is very sensitive to the explosive conditions and forms an interesting tracer, since its short half-life means it is spatially and temporally localized near to its production place and time. Current theoretical models of nova explosions \cite{jose99} predict 
rather small amounts of $^{22}$Na in the ejected shells, below the upper limit derived
by Iyudin et al. \cite{iyudin95}. Nevertheless, such studies have also pointed out significant
uncertainties relating to the nuclear physics input to the model, which affect the critical rates for the $^{21}$Na(p,$\gamma$) and
$^{22}$Na(p,$\gamma$) reactions among others, which are responsible for the 
synthesis (and destruction) of $^{22}$Na,  (see also Ref. \cite{iliadis02} for an extensive study of nuclear
uncertainties in nova nucleosynthesis). 
The uncertainty associated with the $^{21}$Na(p,$\gamma$) reaction rate 
has recently been reduced \cite{bishop03}.
 Reducing the uncertainty in the 
$^{22}$Na(p,$\gamma$) reaction rate may help to constrain nova 
models, improve the estimates on the amount of $^{22}$Na synthesized
during nova outbursts and, hence, dictate the distance at which a $\gamma$-ray flux from $^{22}$Na may be detected.

In the past, several methods have been employed in order to obtain the astrophysical reaction rate for the $^{22}$Na(p,$\gamma$) reaction  \cite{seuthe,steg,schmidt,kubono}. The key to such an analysis is a detailed knowledge of properties such as the excitation energy, spin and parity of levels in the unbound region. The conventional approach to this problem is to study the $^{22}$Na(p,$\gamma$) directly by bombarding a specially prepared radioactive $^{22}$Na target with protons and detecting the $\gamma$ rays following proton capture \cite{seuthe,steg}. 
Using this direct approach, Seuthe {\em et al.} found a series of resonances in the $^{22}$Na(p,$\gamma$) reaction for E$_{p}$=0.17 to 1.29 MeV \cite{seuthe}, while  Stegm\"uller {\em et al.} repeated this study for E$_{p}$ = 0.20 - 0.63 MeV and found a new resonance at E$_{p}$=213 keV \cite{steg}.  These are extremely difficult measurements, however, since not only is there the complication of producing a radioactive target, but there is also the problem of the high background rates in such a hostile experimental environment.
Additional information on the relevant states can be obtained from different reactions such as
 $^{24}$Mg(p,d)  \cite{kubono} and  $^{22}$Na($^{3}$He,d) \cite{schmidt}.
 
In this work, we present a complementary approach to the problem in which particle-unbound states were populated in a heavy-ion fusion evaporation reaction and their subsequent $\gamma$ decay investigated with Gammasphere, a 4$\pi$ high resolution $\gamma$-ray spectrometer consisting of 100 large volume, high purity germanium detectors with absolute efficiency of around 9\% for 1.33 MeV $\gamma$ rays \cite{gammasphere}.

A 10 pnA beam of $^{12}$C was accelerated to 22 MeV by the ATLAS accelerator at Argonne National Laboratory and was incident on a 40 $\mu$g/cm$^{2}$ thick $^{12}$C target. The resulting gamma decay was detected using the Gammasphere array.
The fusion channels observed were single proton, neutron or alpha emission leading to $^{23}$Na, $^{23}$Mg and $^{20}$Na respectively. A $\gamma$-$\gamma$ matrix and a $\gamma$-$\gamma$-$\gamma$ cube were produced and analysed to obtain information on the decay schemes. The construction of the decay schemes was straightforward given the small number of residual nuclei produced and their well-known decay schemes at low excitation energies \cite{toi}. 

Large $\gamma$-ray spectrometers are most commonly optimised for the study of high multiplicity cascades ($\sim$ 20 photons) of relatively low energies ($\sim$ 1 MeV). For the type of study undertaken here, the challenge is to detect cascades both with relatively low multiplicity and consisting of $\gamma$ rays with energies which may be above 10 MeV, meaning that particular attention needs to be paid to both energy and efficiency calibrations. 

In obtaining accurate $\gamma$-ray energies, we have applied a correction for the non-linearity of the array as well as the finite recoil correction for large energy $\gamma$ rays emitted from a light nucleus.
The relevant linearity correction was deduced from source data for 25 standard gamma-rays from the decay of $^{152}$Eu and $^{56}$Co.
This linearity curve was extended to higher energies using data from inelastic scattering of 7.5 MeV protons on a $^{11}$B target. 
This correction for non-linearity was then applied to the $\gamma$ rays observed in the present experiment and comparisons made with the literature values for known $\gamma$ rays \cite{toi}. In cases where two coincident transitions were crossed-over by a third transition, the corrected energy sum was compared and found to agree at the $\sim$ 0.5-1 keV level.

In order to assign a multipolarity to the observed transitions, a matrix was generated of $\gamma$ rays detected at all angles against those detected at 90$^{\circ}$ and a matrix of all $\gamma$ rays against those detected at 32$^{\circ}$ and 37$^{\circ}$. The ratio (R$_{DCO}$) of the intensities of transitions in these two matrices when gating on the `all detector' axis was extracted. This ratio was around 0.9(1) for pure dipole transitions and around 1.7(2) for pure quadrupole transitions. Mixed M1/E2 dipole transitions may have various values depending on the value of the mixing ratio. 
As well as angular correlations, it was also possible to assign the spin/parity of states in $^{23}$Mg, on the basis of their similarity in both energy and decay path to analogue states of well-established spin and parity in $^{23}$Na \cite{toi}, for which extensive additional spectroscopic information was obtained in the present work.

The high energy of many of the $\gamma$ rays observed in the present work implies very short (femtosecond) lifetimes which are readily extracted  using the fractional Doppler shift technique \cite{fds} since it may reasonably be assumed for high-lying, unbound states that the feeding is direct.
Experimental Doppler shifts were obtained by fitting the peak centroids of gamma rays de-exciting states of interest in gamma-ray spectra, without Doppler correction applied, at seven sets of detector angles: 32$^{\circ}$, 50$^{\circ}$, 80$^{\circ}$, 90$^{\circ}$, 100$^{\circ}$, 130$^{\circ}$ and 148$^{\circ}$. Clean identification of the relevant peaks was ensured by using $\gamma$-$\gamma$ coincidence data. The fractional Doppler shifts corresponding to each transition were then obtained by dividing by the maximum Doppler shift expected for the given beam energy and target thickness. 
A model prescription was used to relate these fractional Doppler shifts to the lifetime of the parent state.

\begin{table*}[htb]
\caption{Detailed summary of the properties of states in $^{23}$Mg relevant to the $^{22}$Na(p,$\gamma$) reaction. Proton energies are calculated from the Q value of 7579.6(1.4) keV \cite{audi}. Spins and parities from present work unless otherwise stated.}
\begin{center}
\begin{tabular}{ccccccccccc}
\hline
       E$_{x}$ &   E$_{x}$ &  E$_{p}$ (lab)  & I$_{i}$$^{\pi}$ & I$_{f}$$^{\pi}$ & E$_{\gamma}$  & $\tau$  & R$_{DCO}$ & Branch & $\Gamma_p$ & $\omega\gamma$ \\
 (keV) &        &  (keV)    &        &       & (keV)  & (fs)   &    & (\%) & (meV) & (meV) \\
Endt 1998 \cite{endt}  & present & \\ 
\hline
7622(6)    &  7623.4(9) & 45.8(16) & 9/2$^{+}$ & 5/2$^{+}$ & 7172.5(9) & 4(2) & 1.57(24) & 100 & 1.6$^{+2.2}_{-1.0} \times 10^{-13}$ & 1.1$^{+1.5}_{-0.7} \times 10^{-13}$ \\
7643(10)   &  7646.9(26) & 70.4(30) & 3/2$^{+}$ & 5/2$^{+}$ & 7196.0(26) &  & 0.87(15) & 100 & 2.4$^{+3.4}_{-1.5} \times 10^{-9}$ & 6.8$^{+9.7}_{-4.3} \times 10^{-10}$ \\
           &  7769.2(10) & 198.2(19) & (9/2$^{-}$) & 9/2$^{+}$ & 5054.8(6) & 2(1) & & 58(8) &  & 4.0$_{-4.0}$\\
           &        &       &    & 11/2$^{+}$ & 2316.9(5) &  & & 42(7) \\
7783(3)    &  7779.9(9) & 209.4(17) & (11/2$^{+}$) &7/2$^{+}$ & 5729.1(11) & $<$1 & 1.42(11) & 33(6)  & (6 $\times 10^{-2}$) & (5 $\times 10^{-2})$ \footnote{) estimates assuming l=2, s$_{l}$=0.04})\\
           &            &           &            & 9/2$^{+}$ & 5067.1(11) & & &66(8) \\
7783(3)    &  7784.6(11) & 214.3(18) & (7/2$^{+}$) & 5/2$^{+}$ & 7333.7(11) & 10(3) & 0.89(5) & 100 &       &  1.8$^{+0.7}_{-0.7}$ \footnote{) adopted from Angulo {\em et al.} \cite{angulo}.})\\
7801(2)     &  unobs.     &  231.5(24)   & 5/2$^{+}$ \footnote{) taken from Endt \cite{endt}.})   &           &            &       &         &     &       &  2.2$^{+1.0}_{-1.0}$ \footnote{) taken from Per\"aj\"arvi {\em et al.} \cite{perajarvi}.})\\
7857(2)    &  7851.5(14) & 284.3(20) & (7/2$^{+}$) & 9/2$^{+}$ & 5138.1(13) &     &         & 100 &       &  15.8$^{+3.4}_{-3.4}$ \footnotemark[2]) \\
8017(2)    &  8015.3(17) & 455.5(23) &  (5/2$^{+}$-11/2$^{+}$)  & 9/2$^{+}$ & 5300.2(9) & & & 71(16) &    &  68$^{+20}_{-20}$ \footnotemark[2]) \\
           &             &           &                          & 7/2$^{+}$ & 5966.7(11) & &  &29(12) \\
8166(2)    & 8159.7(20) &  606.5(25)      &5/2$^{+}$ \footnotemark[3]) & 7/2$^{+}$ & 6109.5(18) &  &   & 100 &  &235$^{+33}_{-33}$\footnotemark[2]) \\
\hline
\end{tabular}
\end{center}
\label{22natable}
\end{table*}


The Gammasphere array affords the possibility of determining the energy of the resonances to superior accuracy than that obtainable with a spectrometer. It also allows their decay path to be observed and through angular distribution measurements, the spin/parity of these resonances may be inferred or at least restricted to some plausible range. The $\gamma$-ray energies, angular correlation ratios, branching ratios and lifetimes of proton-unbound states in $^{23}$Mg are presented in table\ \ref{22natable}. The information contained in the table is only a very limited selection from the detailed spectroscopy carried out on both $^{23}$Na and $^{23}$Mg in the present work, which will be fully reported in \cite{dgj}.
 Our observations and deductions of the properties of the relevant states are broadly compatible with those made previously with a few notable exceptions summarised below.

We have been able to make firm spin assignments for the previously reported 7623 and 7647 keV levels. In an earlier $^{24}$Mg(p,d) study, the angular distribution of the particle groups corresponding to the 7623 keV state were assigned a multipolarity L=4, i.e. the available spin possibilities are 7/2$^{+}$ and 9/2$^{+}$ \cite{kubono}. We find that the 7173 keV $\gamma$ ray which depopulates the 7623 keV state has an angular correlation ratio consistent with an E2 assignment. Since the 7173 keV $\gamma$ ray feeds a known 5/2$^{+}$ level, we dismiss the 7/2$^{+}$ possibility and assign a spin/parity of 9/2$^{+}$ to the 7623 keV level. Similar arguments firmly fix the 7647 keV level as J$^{\pi}$=3/2$^{+}$.

\begin{figure}[htb]
\includegraphics[width=\linewidth, clip=]{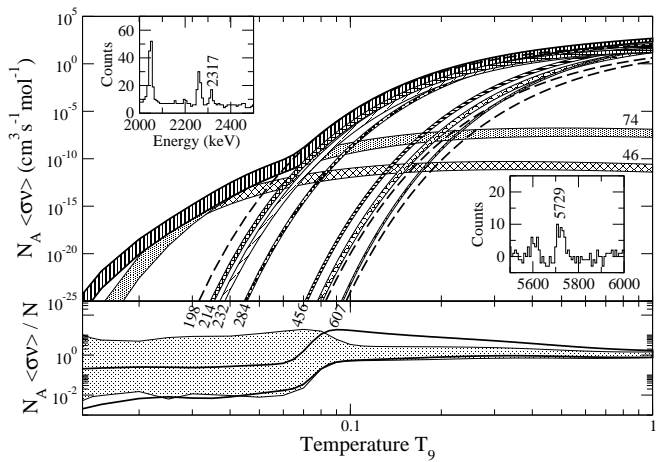}
\caption{(Top panel) Contribution of individual resonances to the total reaction rate for the $^{22}$Na(p,$\gamma$) reaction. The contributions are labeled with E$_{p}$ (lab) in keV. (Inset top left) Spectrum from $\gamma$-$\gamma$-$\gamma$ cube gated on 2739 and 1600 keV transitions. (Inset bottom right) Spectrum from $\gamma$-$\gamma$-$\gamma$ cube gated on 450 and 1600 keV transitions. (Bottom Panel) Comparison of the total reaction rates normalised to the analytical rate presented by Angulo {\em et al.} \cite{angulo}. The shaded region is the uncertainty in the rates given in the Angulo compilation \cite{angulo}. The solid lines bound the uncertainties in the present work.}
\label{22narate}       
\end{figure}

 We find a state at 7784.6(11) keV which we believe corresponds to the state for which Stegm\"uller {\em et al.} adopt an excitation energy of 7783.4(22) \cite{steg} and which we also find to decay solely to the 450 keV 5/2$^{+}$ level. We find, however, a previously unreported nearby state at 7779.9(9) keV which decays to the 7/2$^{+}$ and 9/2$^{+}$ states in the yrast band (see bottom right inset to Fig.~\ref{22narate}), and a further state at 7769.2(10) keV which decays both to the lowest 9/2$^{+}$ and 11/2$^{+}$ states (see top left inset to Fig~\ref{22narate}). We associate this state by mirror symmetry, on the basis of its energy and decay branching, with a 9/2$^{-}$ state in $^{23}$Na which is part of the (1/2$^{-}$)$_{2}$ band (a full discussion of the mirror symmetry of high spin states in $^{23}$Na and $^{23}$Mg will be reported elsewhere \cite{dgj}). 

In evaluating the total reaction rate, we have used the results of the present work in conjunction with directly-measured resonance strengths from the work of Seuthe {\em et al.} \cite{seuthe} and Stegm\"uller {\em et al.}. Excitation energies and strengths of resonances at 7583, 8061, 8074 and 8153 keV which were unobserved in the present work were taken from Angulo {\em et al.} \cite{angulo} and that of the 7801 keV IAS resonance from Per\"aj\"arvi {\em et al.} \cite{perajarvi}. The evaluation was truncated at the 8160 keV resonance.
Rates for the 7623 keV and 7647 keV resonances were evaluated using proton spectroscopic factors obtained from a $^{22}$Na($^{3}$He,d) study \cite{schmidt}, in conjunction with the more accurate resonance energies and the firm spin/parities obtained in the present work.
Assessing the possible contribution of the newly-discovered 7769 keV state is problematic, since if it has J$^{\pi}$ = 9/2$^{-}$, it would make a non-negligible contribution to the overall reaction rate. This contribution is constrained, however, by a single measured upper limit on resonance yield for the $^{22}$Na(p,$\gamma$) reaction for E$_{p}$ (lab) = 202 keV. This constraint is complicated by the fact that as much as 50\% of the 7769 keV resonance decays by a $\gamma$ ray with an energy less than 3 MeV, which would have been undetectable by the apparatus used in this direct measurement \cite{steg}. Given also the uncertainty in resonance energy,  we set a conservative upper limit on the resonance strength for the 7769 keV state of 4.0 meV. We have not explicitly included a contribution from the newly discovered 7780 keV state because it is most likely high spin (11/2$^{+}$), and its contribution is subsumed within the earlier direct measurement of the neighbouring 7783 keV resonance \cite{steg}. Direct capture has been neglected as it has been shown to have a negligible contribution to the total reaction rate \cite{seuthe,steg}. Our re-evaluation of the reaction rate is presented in figure~\ref{22narate}.

There are clearly several important differences between the present reaction rate and those presented in earlier work \cite{steg,schmidt}. First, by fixing the energies of the 7623 and 7647 keV resonances with the superior resolution of Germanium detectors and by assigning these resonances firm spin/parities of 9/2$^{+}$ and 3/2$^{+}$, respectively, we have been able to significantly reduce the uncertainty in the reaction rate at the lowest temperatures.
Second, at temperatures around T$_{9}$ = 0.1, typical of nova outbursts, we find that the 7769 keV resonance (E$_{p} = 198$ keV), which was not included in earlier evaluations of the reaction rate, may make a substantial contribution. A new direct measurement of the energy region, E$_{p}$ (lab) = 180 - 220 keV is, therefore, clearly warranted.

We have performed an analysis of the impact of the new upper and lower limits on the rate for the
$^{22}$Na(p,$\gamma$) reaction on the amount of $^{22}$Na ejected
during nova outbursts. Two evolutionary sequences for a nova outburst hosting an ONe 
white dwarf of 1.25 \msun, have been computed by means of a spherically symmetric,
hydrodynamic, implicit code, in Lagrangian formulation, extensively used in the modeling
of such explosions (see Ref. \cite{josehernanz98} for details). A mean mass fraction of 
$3.3 \times 10^{-4}$ for $^{22}$Na \cite{jose99} was obtained with the previous rate of  Stegm\"uller {\em et al}. \cite{steg} . Using the new upper and lower limits, we obtain values of $1.2 \times 10^{-4}$ and $3.9 \times 10^{-4}$, respectively. The former value implies a reduction by a factor $\sim 3$, or roughly a factor of 1.7 in terms of the maximum detectability
distance for the 1.275 MeV $\gamma$ ray. Using the previous rate estimate, 
 the maximum detectability distance with the INTEGRAL spectrometer SPI, was around 1 kiloparsec \cite{hernanz2002,gomez98}. If the reaction rate were close to the present upper limit, then the distance would reduce to $\sim$ 0.6 kiloparsecs, 
significantly reducing the chances of detecting the characteristic $^{22}$Na $\gamma$ ray in nearby novae.

The detailed spectroscopic information presented in the present work from which this reevaluation has been made, was obtained by employing a large Germanium detector array in an unconventional fashion. New states were identified in the relevant excitation region; the accuracy of the excitation energy of the previously known states was improved and several ambiguities in spin/parity assignment were resolved. It appears that this method may be quite general and may be applicable to many similar reaction rate determinations.

This work was supported in part by the U.S. Department of Energy, Nuclear Physics Division, under contract No. W-31-109-ENG38, and by Spanish MCYT Grant AYA2002-04094-C03-03.


\end{document}